\documentstyle[12pt]{article}

\def\thebibliography#1{\begin{center}
                          \underline{REFERENCES}
                       \end{center}\list
 {[\arabic{enumi}]}{\settowidth\labelwidth{[#1]}\leftmargin\labelwidth
 \advance\leftmargin\labelsep
 \usecounter{enumi}}
 \def\newblock{\hskip .11em plus .33em minus .07em}
 \sloppy\clubpenalty4000\widowpenalty4000
 \sfcode`\.=1000\relax}

\begin{document}
\bibliographystyle{prsty}

$\mbox{  }$\vskip 4cm
\begin{center}

A GAUGE INVARIANT UNITARY THEORY FOR PION PHOTOPRODUCTION
\footnote{Presented at the 14th European Conference on Few-Body Problems in
Physics}
\vskip \baselineskip
\vskip \baselineskip
\vskip \baselineskip
C.~H.~M.~van~Antwerpen and I.~R.~Afnan
\\ School of Physical~Sciences, The Flinders~University of South~Australia, \\
Bedford~Park, South~Australia~5042, Australia
\end{center}
\vskip \baselineskip
\vskip \baselineskip
\noindent\underline{ABSTRACT}:
A covariant, unitary and gauge invariant theory for pion photoproduction on a
single nucleon is presented. To achieve gauge invariance at the operator
level one needs to include both the $\pi N$ and $\gamma\pi N$ thresholds.
The final amplitude can be written in terms of a distorted wave in the final
$\pi N$ channel provided one includes additional diagrams to the standard
Born term in which the photon is coupled to the final state pion and nucleon.
These additional diagrams are required in order to satisfy gauge invariance.
\vskip \baselineskip
\vskip \baselineskip
Most calculations to date for pion photoproduction on a single nucleon have
included two ingredients: (i) A Born term which is taken to be gauge
invariant. (ii) A final state interaction or distortion that is needed to
satisfy the Watson theorem or unitarity.  However, the inclusion of the
pionic degrees of freedom into the problem changes the current and charge
distribution and thus the coupling of the photon to the nucleon. Hence
any consistent  theory of pion photoproduction has to include the effect of
the pionic degrees of freedom on the charge and current distribution.
In other words one needs to satisfy unitarity and gauge invariance at the
same time. Clearly, the inclusion of both of these symmetries to all orders
requires a full solution to the field theory. Here, we propose to present a
formulation that satisfies two-body unitarity and gauge invariance
to first order in the electromagnetic coupling.

The starting point of this formulation is the one particle irreducible $\pi NN$
three-point Green's function given by
\begin{eqnarray}
   G(q,p';p) &=& \int\,d^4x_1d^4x_2d^4x_3\,
                      e^{i(p'\cdot  x_3+q\cdot x_1-p\cdot x_2)}\,
             <0|T(\phi(x_1)\psi(x_2)\bar{\psi}(x_3))|0> \nonumber \\
   &=& S(p')\,\Delta(q)\,\Lambda_5^{(1)\dagger}(q,p';p)\,S(p)\ ,
                                                 \label{eq:1}
\end{eqnarray}
where the spin, isospin labels have been
suppressed. Here, $S$ is the nucleon propagator, $\Delta$ is the
pion propagator, and $\Lambda_5^{(1)\dagger}$ is the one pion irreducible
$\pi NN$ vertex. Since at this stage we are considering two-body
unitarity only, we will restrict the dressing of the nucleon so that
the final $\pi N\leftarrow\gamma N$ amplitude has at most one pion in
every intermediate state. This allows us to write the $\pi NN$
three-point Green's function as
\begin{equation}
   G(q,p';p)=S_0(p')\,\Delta(q)\,
                 \Lambda_5^{(1)\dagger}(q,p';p)\,S(p)\ ,
                                                      \label{eq:2}
\end{equation}
where $S_0(p) = (\not{\!\! p} - m)^{-1}$
and $S(p)=(\not{\!\! p} - m -\Sigma^{(1)}(\not{\!\! p}))^{-1}$, with
$\Sigma^{(1)}$ including all contributions to two-body unitarity from
mass renormalization.

The photoproduction amplitude is then constructed by gauging
Eq.~(\ref{eq:2}) and applying the LSZ reduction to the resulting
four-point Green's function. The method of gauging employed was the
minimal substitution $p_\mu\rightarrow p_\mu - eA_\mu$,
where $p_\mu$ is the four momentum of the particle. This results in the
following substitutions~\cite{Ka59,Oh89}
\begin{eqnarray}
      S(\hat{p})&\rightarrow& S(\hat{p})
   + S(p') \Gamma_{\mu}(k,p',p) S(p)A^\mu \quad , \label{eq:S} \\
   \Delta(\hat{q})&\rightarrow& \Delta(\hat{q}) +
              \Delta(q') \Gamma^{\pi}_{\mu}(k,q',q) \Delta(q)A^\mu \quad ,
              \label{eq:D} \\
   \Lambda^{(1)\dagger}_5(\hat{q},\hat{p}',\hat{p})
   &\rightarrow& \Lambda^{(1)\dagger}_5(\hat{q},\hat{p}',\hat{p}) +
   \Gamma_{\mu}^{CT}(k,q,p',p)A^\mu \quad . \label{eq:pBB}
\end{eqnarray}
i.e.\ the photon couples to all possible external lines and vertices present.
We have followed a procedure developed by Ohta~\cite{Oh89} in which
(i)~a Taylor series expansion is assumed to exist for the form factors present,
(ii)~perturbation theory can be used to replace various operators by
eigenvalues, and
(iii)~we restrict ourselves to first order in the electromagnetic coupling.
Applying the substitutions of Eqs.~(\ref{eq:S}-\ref{eq:pBB}) to
Eq.~(\ref{eq:2}) leads to the four classes of diagrams~\cite{NK90} describing
the photoproduction amplitude,
\begin{equation}
   \Lambda_5^{(1)\dagger}\,S\,\Gamma_\mu^{(1)}
       + \Gamma_\mu^{(2)}\,S_0\,\Lambda_5^{(1)\dagger} +
       \Gamma_\mu^{\pi(2)}\,\Delta\,\Lambda_5^{(1)\dagger} +
       \Gamma_\mu^{CT(1)}\ ,
                                                        \label{eq:3}
\end{equation}
where $\Gamma_\mu^{(1)}$ is the one-particle irreducible photon
nucleon vertex, $\Gamma_\mu^{\pi(2)}$ is the photon pion vertex which
is taken to be the bare vertex, and $\Gamma_\mu^{CT(1)}$ is the $\pi
N\leftarrow\gamma N$ amplitude resulting from the coupling of the
photon to the $\pi NN$ vertex, $\Lambda_5^{(1)\dagger}$.  Here the
irreducibility is given with respect to the number of pions and
nucleons only.  This amplitude is by definition gauge invariant with
the photon vertices satisfying their corresponding Ward-Takahashi
identities~\cite{Ka59,Wa50,Ta57}.

To establish the fact that the amplitude for $\pi N\leftarrow\gamma
N$, as given in Eq.~(\ref{eq:3}) satisfies two-body unitarity, we
follow the procedure of Araki and Afnan~(AA)~\cite{AA87} and
classify a diagram's contribution to the amplitude according to
it's irreducibility. However, unlike AA who included first the $\pi
N$ threshold for two-body unitarity and then the $\pi\pi N$ and
$\gamma\pi N$ cuts for three-body unitarity, in this case we include
both the $\pi N$ and $\gamma\pi N$ unitarity cuts, since the
corresponding branch points are at the same energy, to satisfy
two-body unitarity.

Exposing the $\pi N$ unitarity cut results in the following integral
equation for the $\pi N$ amplitude~\cite{AA87}
\begin{equation}
   t^{(0)}=v(1+gt^{(0)}) \label{eq:t0i}
\end{equation}
where $v$ is the Born amplitude given by
$v=t^{(2)}+\Lambda^{(2)\dagger}_5 S_{0}\Lambda^{(2)}_5$ ,
and $g=S_0\,\Delta$ is the $\pi N$ propagator.
For the $\pi N\leftarrow\gamma N$ amplitude exposing the $\pi N$ cut gives
\begin{equation}
   M^{(0)}=\tilde{v}+vgM^{(0)}\quad ,
   \label{eq:M0}
\end{equation}
with $\tilde{v}=M^{(2)}+\Lambda^{(2)\dagger}_5 S_{0}\Gamma^{(2)}$.
In this case $\tilde{v}$ is not the Born amplitude, since exposing the
$\gamma\pi N$ cut in $\Gamma^{(2)}$ gives
\begin{equation}
   \Gamma^{(2)}=\Gamma^{(3)}+\Gamma^{3(3)}g\Lambda^{(1)\dagger}_5
   \quad ,
   \label{eq:f2g2}
\end{equation}
where $\Gamma^{3(3)}$ is the $N\leftarrow\gamma\pi N$ amplitude.
Similarly, exposing the $\gamma\pi N$ cut in $M^{(2)}$ gives
\begin{equation}
   M^{(2)}=\Gamma^{CT(3)}+\tilde{F}^{(3)}_{3;c}g\Lambda^{(1)\dagger}_5
                  +\Gamma^{\pi(2)}\Delta\Lambda^{(1)\dagger}_5
                  +\Gamma^{(3)}S_{0}\Lambda^{(1)\dagger}_5
                  \quad , \label{eq:tt2}
\end{equation}
where $\Gamma^{CT(3)}$ is the $\pi N\leftarrow\gamma N$ amplitude,
$\tilde{F}^{(3)}_{3;c}$ is the $N\pi\leftarrow\gamma\pi N$ amplitude, and
$\Gamma^{\pi(2)}$ is the $\pi\leftarrow\gamma\pi$ amplitude.
This results in $\tilde{v}$ being given by
\begin{equation}
   \tilde{v}=\Gamma^{CT(3)}+\tilde{F}^{(3)}_{3;c}g\Lambda^{(1)\dagger}_5
            +\Gamma^{\pi(2)}\Delta\Lambda^{(1)\dagger}_5
            +\Gamma^{(3)}S_{0}\Lambda^{(1)\dagger}_5 \nonumber \\
            +\Lambda^{(2)\dagger}_5 S_{0}\Gamma^{(3)}
            +\Lambda^{(2)\dagger}_5
              S_{0}\Gamma^{3(3)}g\Lambda^{(1)\dagger}_5
            \label{eq:ttv}
\end{equation}
which contains more physics than just the Born amplitude.  By comparing
the above result for the photoproduction amplitude which satisfies two-body
unitarity, with the amplitude in Eq.~(\ref{eq:3}), we can establish that
they are in fact identical~\cite{AA93}.

If we now iterate Eq.~(\ref{eq:M0}) and make use of Eq.~(\ref{eq:t0i}), we find
that
\begin{equation}
   M^{(0)}=\bigl(t^{(0)}g+1\bigr)\tilde{v}
   =\bigl(t^{(0)}g+1\bigr)\tilde{v}_B+\bigl(t^{(0)}g+1\bigr)\tilde{v}_R
   \quad , \label{eq:litm}
\end{equation}
where $\tilde{v}_B$ is the Born amplitude and $\tilde{v}_R$ contains the
additional diagrams required
to maintain gauge invariance which are illustrated in Fig.~\ref{fig:rest}.

\begin{figure}
   \vskip 3.5 cm
   \hskip 0 cm
   \caption{These are the non-Born diagrams which contribute to the pion
            photoproduction amplitude.  The numbers in the circle give
            the irreduciblity of each amplitude.}
   \label{fig:rest}
\end{figure}

This result also proves that the commonly used procedure for
unitarizing the Born amplitude in terms of the $\pi N$ amplitude,
i.e.\
\begin{equation}
   M^{(0)} = (t^{(0)}\,g + 1)\,\tilde{v}_{B}
   \quad , \label{eq:7}
\end{equation}
does not satisfy gauge invariance. This is due to the fact that the
derivation of Eq.~(\ref{eq:7}) involves the inclusion of the $\pi N$
threshold, which is what unitarity requires, and only the lowest
order diagrams containing a $\gamma\pi N$ intermediate state, into
$\tilde{v}_{B}$. However, to satisfy gauge invariance, the full
$\gamma\pi N$ threshold should also be included, which is what one
expects considering the fact that the two thresholds start at the
same energy.  We should also note that the additional terms resulting from the
inclusion of the $\gamma\pi N$ threshold give rise to the dressing
of the vertices in $\tilde{v}_{B}$, and this dressing is required
to satisfy gauge invariance.

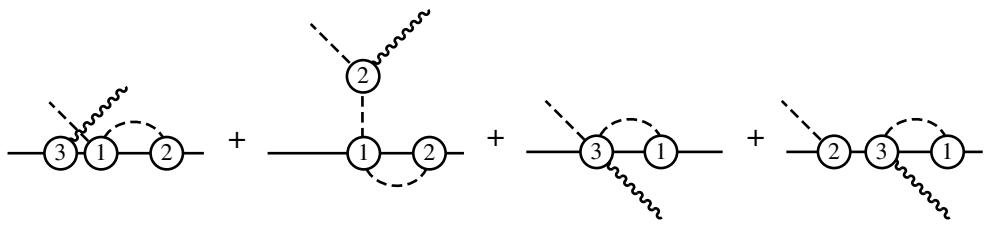
\end{document}